\documentclass{iopart}
\usepackage{amsfonts}
\eqnobysec
\def\sgn{\mathop{\rm sgn}\nolimits}
\def\res{\mathop{\rm res}}
\def\mod{\mathop{\rm mod}\nolimits}
\def\erf{\mathop{\rm erf}\nolimits}

\begin{document}

\title[Towards Inverse Scattering for non decaying KPII]{Towards an Inverse Scattering theory for non decaying potentials of the heat
equation\footnote[3]{
Work supported in part by INTAS 99-1782, by Russian Foundation for Basic
Research 99-01-00151 and by PRIN 97 `Sintesi'.}}

\author{M Boiti\dag, F Pempinelli\dag, A K Pogrebkov\ddag and B Prinari\dag}
\address{\dag\ Dipartimento di Fisica dell'Universit\`{a} and Sezione INFN,
73100 Lecce, Italy}
\address{\ddag\ Steklov Mathematical Institute Moscow, 117966, GSP-1, Russia}

\begin{abstract}
The resolvent approach is applied to the spectral analysis of the heat
equation with non decaying potentials. The special case of potentials with
spectral data obtained by a rational similarity transformation of the
spectral data of a generic decaying potential is considered. It is shown
that these potentials describe $N$ solitons superimposed by B\"{a}cklund
transformations to a generic background. Dressing operators and Jost
solutions are constructed by solving a $\overline{\partial }$-problem
explicitly in terms of the corresponding objects associated to the original
potential. Regularity conditions of the potential in the cases $N=1$ and $
N=2 $ are investigated in details. The singularities of the resolvent for
the case $N=1$ are studied, opening the way to a correct definition of the
spectral data for a generically perturbed soliton.
\end{abstract}

\section{Introduction}

The operator
\begin{equation}
\mathcal{L}(x,\partial _{x}^{})=-\partial _{x_{2}}^{}+\partial _{x_{1}}^{2}-u(x),\quad \quad
x=(x_{1},x_{2})  \label{1}
\end{equation}
which defines the well-known equation of heat conduction, or heat equation
for short, from the beginning of the seventies \cite{dryuma74,zakharov74} is
known to be associated to the Kadomtsev--Petviashvili equation in its
version called KPII
\begin{equation}
(u_{t}^{}-6uu_{x_{1}}^{}+u_{x_{1}x_{1}x_{1}}^{})_{x_{1}}^{}=-3u_{x_{2}x_{2}}^{}. \label{KPI}
\end{equation}
The spectral theory for the equation of heat conduction with real potential $
u(x)$ was developed in \cite{BarYacoov}--\cite{Grinevich}, but only the case
of potentials rapidly decaying at large distances in the $x$-plane was
considered. We are interested in including in the theory potentials with
one-dimensional asymptotic behaviour and in particular potentials describing
$N$ solitons on a generic background. In trying to solve the analogous
problem for the nonstationary Schr\"{o}dinger operator, associated to the
KPI equation, a new general approach to the inverse scattering theory was
introduced, which was called resolvent approach. See \cite{total}--\cite
{towards} for the nonstationary Schr\"{o}dinger operator and \cite{GP} for
the Klein--Gordon operator. Some results for the operator (\ref{1}) were
given in \cite{PhysicaD}. For the specific case of $N$ solitons on a
background for the KPI equation see especially \cite{towards,Steklov}.

Here we apply the same approach to the heat equation and construct a potential $u'(x)$
describing $N$ solitons superimposed to a generic background potential $u(x)$. Superimposition
is performed by means of a rational similarity transformation of the spectral data of $u(x)$.
This procedure supplies us not only with the B\"{a}cklund transformation of the potential
$u(x)$, but also with the corresponding Darboux transformations of the Jost solutions and with
the spectral theory for the transformed potential $u'(x)$. All related mathematical entities
such as the extended resolvent $M'$, the dressing and dual dressing operators $ \nu'$ and
$\omega '$ and the Jost and dual Jost solutions $\Phi '$ and $\Psi '$ corresponding to $u'(x)$
are given explicitly in terms of the same objects associated to the background potential
$u(x)$. In \cite{KPIIproceedings,Barbara} some preliminary results were presented for $N=1$
and 2 by using recursively the binary Darboux transformations.

We show that the main mathematical object of the theory, i.e.\ the extended resolvent $M'$, is
given as a sum of two terms. The first one is obtained by dressing with the operators $\nu '$
and $\omega '$ the resolvent $M_{0}$ of the bare heat operator $\mathcal{L}^{}_{0}
(x,\partial _{x}^{})=-\partial _{x_{2}}^{}+\partial _{x_{1}}^{2}$, while the second one,
$m'$, takes into account the discrete part of the spectrum. The dressing operators $\nu '$ and
$\omega '$ are constructed by solving a $\overline{\partial }$-problem for the transformed
spectral data.

The theory with respect to the nonstationary Schr\"{o}dinger equation is in some respects
simpler and in some other respects unexpectedly more difficult. We give the explicit
expression for $\Phi '$, $\Psi '$, and $u'$ for any $N$, but the reality and regularity
conditions for the potential are rather involved and we examine in details only the cases
$N=1,2$. We study the singularities of the resolvent in the case $N=1$. This resolvent can be
used for investigation of the spectral theory of operator (\ref{1}) with the potential being a
perturbation of the potential $u'(x)$ obtained by adding to it a `small' function $ u_{2}\left(
x\right) $ rapidly decaying on the $x$-plane. It can be shown that under such a perturbation
of the potential the Jost solutions get singularities more complicated than poles, but on the
other side they have no additional cuts on the complex plane of spectral parameter, in
contrast with the nonstationary Schr\"{o}dinger case. This means, however, that also in the
case of the perturbed heat equation the standard definition of spectral data for a generic non
decaying potential must be modified. The solution of this problem is deferred to a future
work.

\section{Direct and inverse problems in the case of rapidly decaying
potentials}

In the framework of the resolvent approach we work in the space $\mathcal{S}'$ of tempered
distributions $A(x,x';q)$ of the six real variables $x$, $x'$, $q\in \mathbb{R}^{2}$. It is
convenient to consider $q$ as the imaginary part of a two-dimensional complex variable $
\mathbf{q=q}_{\Re }+\rmi \mathbf{q}_{\Im }=(\mathbf{q}_{1},\mathbf{q}_{2})\in \mathbb{C}^{2}$
and to introduce the ``shifted'' Fourier transform
\begin{equation}
A(p;\mathbf{q})=\frac{1}{(2\pi )^{2}}\int \!\!\rmd x \!\!\int \!\!\rmd x'\,\rme_{}^{\rmi(p+
\mathbf{q}_{\Re })x-\rmi \mathbf{q}_{\Re }x'} A(x,x';\mathbf{q}_{\Im })  \label{2a}
\end{equation}
where $p\in \mathbb{R}^{2}$, $px=p_{1}x_{1}+p_{2}x_{2}$ and $\mathbf{q}_{\Re }x=$
$\mathbf{q}_{1\Re }x_{1}+\mathbf{q}_{2\Re }x_{2}$. We consider the distributions $A(x,x';q)$
and $A(p;\mathbf{q})$ as kernels in two different representations, the $x$-representation and
the ($p,\mathbf{q}$)-representation, respectively, of the operator $A(q)$ ($A$ for short). The
composition law in the $x$-representation is defined in the standard way, that is
\begin{equation}
(AB)(x,x';q)=\!\!\int \!\!\rmd x''\,A(x,x'';q)\,B(x'',x';q). \label{180xspace}
\end{equation}
Since the kernels are distributions this composition is neither necessarily
defined for all pairs of operators nor associative. In terms of the
($p,\mathbf{q}$)-representation (\ref{2a}) this composition law is given by a
sort of a ``shifted'' convolution,
\begin{equation}
(AB)(p;\mathbf{q})=\!\!\int \!\!\rmd p'A(p-p';\mathbf{q}+p')B(p';\mathbf{q}). \label{180}
\end{equation}
On the space of these operators we define the conjugation $A_{}^{\ast }$ and the shift
$A_{}^{(\mathbf{s})}$ for the complex parameter $\mathbf{s} \in \mathbb{C}^{2}$, that in the
$x$- and ($p,\mathbf{q}$)-representations read, respectively,
\begin{equation}
A_{}^{\ast }(x,x';q)=\overline{A(x,x';q)},\qquad A_{}^{\ast
}(p;\mathbf{q})=\overline{A(-p;-\overline{\mathbf{q}})} \label{3a}
\end{equation}
where bar denotes complex conjugation, and
\begin{equation}
\fl A_{}^{(\mathbf{s})}(p;\mathbf{q})=A(p;\mathbf{q+s}),\qquad A_{}^{(
\mathbf{s})}(x,x';q)=\rme_{}^{\rmi \mathbf{s}_{\Re }(x-x')}A(x,x';q+\mathbf{s}_{\Im }).
\label{2a10}
\end{equation}
For any operator $A$, thanks to the fact that its kernel belongs to the space
$\mathcal{S}'(\mathbb{R}^{6})$, we can consider its $\bar{
\partial}_{j}$-differentiation (in the sense of distributions) with respect
to the complex variables $\mathbf{q}_{j}$ ($j=1,2$):
\begin{equation}
\fl (\bar{\partial}_{j}^{}A)(p;\mathbf{q})=\frac{\partial A(p;\mathbf{q})}{
\partial \overline{\mathbf{q}}_{j}^{}},\qquad
(\bar{\partial}_{j}^{}A)(x,x';q)=\frac{\rmi }{2} \Bigl(x_{j}^{}-x_{j}'+\frac{\partial
}{\partial q_{j}^{}} \Bigr)A(x,x';q),  \label{2a12}
\end{equation}
where the formula for the kernel of $\bar{\partial}_{j}A$ in the
$x$-representation is obtained by (\ref{2a}).

The set of differential operators is embedded in the introduced space of
operators by means of the following extension procedure. Any given
differential operator $\mathcal{A}(x,\partial _{x})$ with kernel
\begin{equation}
A(x,x')=\mathcal{A}(x,\partial _{x}^{})\delta (x-x'), \label{kern}
\end{equation}
$\delta (x-x')=\delta (x_{1}-x_{1}')\delta (x_{2}-x_{2}')$ being the two-dimensional $\delta
$-function, is replaced by the operator $A(q)$ with kernel
\begin{equation}
A(x,x';q)\equiv \rme_{}^{-q(x-x')}A(x,x')=\mathcal{ A}(x,\partial _{x}^{}+q)\delta (x-x').
\label{1'}
\end{equation}
We refer to the operator $A(q)$ constructed in this way as the ``extended''
version of the differential operator $\mathcal{A}$. It is easy to see that
in terms of the ($p,\mathbf{q}$)-representation (\ref{2a}) the dependence on
the $\mathbf{q}$-variables of the kernels of these extensions of the
differential operators is polynomial. In~particular, let $D_{j}$ denote the
extension of the differential operator $\partial _{x_{j}}$ ($j=1,2$), i.e.
according to~(\ref{1'})
\begin{equation}
D_{j}^{}(x,x';q)=(\partial _{x_{j}}^{}+q_{j}^{})\delta (x-x'),\qquad j=1,2, \label{10}
\end{equation}
then $D_{j}$ in the ($p,\mathbf{q}$)-representation takes the form
\begin{equation}
D_{j}(p;\mathbf{q})=-\rmi \mathbf{q}_{j}\delta \left( p\right) .
\end{equation}

An operator $A$ can have an inverse in terms of the composition law (\ref {180xspace}),
(\ref{180}), say $AA_{}^{-1}=I$ (in general left and right inverse can be different), where
$I$ is the unity operator,
\begin{equation}
I(x,x';q)=\delta (x-x'),\qquad I(p;\mathbf{q})=\delta (p). \label{15}
\end{equation}
In order to make the inversion $A^{-1}$ of an (extended) differential
operator $A$ uniquely defined we impose the condition that the product $
(A^{-1})^{(\mathbf{s})}A^{-1}$ exists and is a bounded function of $\mathbf{s
}$ in a neighborhood of $\mathbf{s}=0$. Let us consider as an example the
operator $D_{1}-a$, where, being $a$ a complex constant, we write for
shortness $a$ instead of $aI$, according to a general notation we use in the
following. For its inverse operator we get
\begin{equation}
\fl (D_{1}^{}-a)_{}^{-1}(x,x';q)=\sgn (q_{1}^{}-a_{\Re
}^{})\rme_{}^{(a-q_{1})(x_{1}-x_{1}')}\theta ((q_{1}^{}-a_{\Re }^{})(x_{1}^{}-x_{1}'))\delta
(x_{2}^{}-x_{2}'),  \label{101}
\end{equation}
that is just the standard resolvent of the operator $\partial _{x_{1}}$. In
terms of the ($p,\mathbf{q}$)-representation this inverse operator is given
by
\begin{equation}
(D_{1}-a)^{-1}\left( p;\mathbf{q}\right) =\rmi (\mathbf{q}_{1}-\rmi a)^{-1}\delta
(p).  \label{Dpq}
\end{equation}
The simplicity of this formula makes clear the usefulness of the ($p,\mathbf{
q}$)-representation. Notice also that in this representation the necessity of requiring the
boundedness condition introduced above for the inverse is especially evident since it is just
this condition that excludes the presence of additional terms of the type $\delta
(\mathbf{q}_{1}^{}-\rmi a)$ in (\ref{Dpq}).

According to this general construction the extended operator $L(q)$
corresponding to the operator (\ref{1}) is given by
\begin{equation}
L=L_{0}^{}-U,  \label{12}
\end{equation}
where $L_{0}$ $=-D_{2}^{}+D_{1}^{2}$, i.e.\ the extension of $\mathcal{L} (x,\partial _{x})$
in the case of zero potential with kernels
\begin{equation}
\fl L_{0}^{}(x,x';q)=\left[ -\left( \partial _{x_{2}}^{}+q_{2}\right) +\left(
\partial _{x_{1}}^{}+q_{1}\right) ^{2} \right] \delta (x-x'),\quad
L_{0}^{}(p;\mathbf{q})= (\rmi \mathbf{q}_{2}^{}-\mathbf{q}_{1}^{2})\delta (p)  \label{13}
\end{equation}
and $U$ can be called the potential operator since has kernels
\begin{equation}
U(x,x';q)=u(x)\delta (x-x'),\qquad U(p;\mathbf{q})=v(p), \label{a11}
\end{equation}
where $v(p)=\left( 2\pi \right) ^{-2}\int \!\!\rmd x\,\rme^{\rmi px}u(x)$ is the Fourier
transform of the potential $u(x)$. Below we always suppose that $u(x)$ is
real, which by (\ref{3a}) is equivalent to
\begin{equation}
L_{}^{\ast }=L.  \label{16}
\end{equation}

The main object of our approach is the (extended) resolvent $M(q)$ of the
operator $L(q)$, which is defined as the inverse of the operator $L$, that
is
\begin{equation}
LM=ML=I.  \label{14}
\end{equation}
In order to ensure the uniqueness of $M$ we require that the product $M^{(
\mathbf{s})}M$ is a bounded function in the neighborhood of $\mathbf{s}=0$.
Then, in particular, the resolvent $M_{0}$ of the bare operator $L_{0}$ has
in the ($p,\mathbf{q}$)-representation kernel
\begin{equation}
M_{0}^{}(p;\mathbf{q})=\frac{\delta (p)}{\rmi \mathbf{q}_{2}^{}- \mathbf{q}_{1}^{2}}.
\label{251}
\end{equation}
In the $x$-representation we get
\begin{equation}
\fl M^{}_{0}(x,x';q)=\frac{\rme_{}^{-q(x-x')}}{2\pi }\!\!\int\!\! \rmd\alpha \,\bigl[\theta
(q_{1}^{2}-q_{2}^{}-\alpha _{}^{2})-\theta (x_{2}^{}-x_{2}')\bigr]\,\rme_{}^{-\rmi \ell
(\alpha +\rmi q_{1})(x-x')}.  \label{M0xxq}
\end{equation}
The resolvent of $L$ can also be defined as the solution of the integral
equations
\begin{equation}
M=M_{0}^{}+M_{0}^{}UM,\qquad M=M_{0}^{}+MUM_{0}^{}. \label{integralM}
\end{equation}
Under a small norm assumption for the potential we expect that the solution $
M$ is unique (the same for both integral equations) and that it satisfies
the boundedness condition at $\mathbf{s}=0$ for $M^{(\mathbf{s})}M$.

The resolvent is directly connected with the Green's functions of the operator (\ref{1}).
Indeed, since by definition (\ref{1'}) the product $ \rme^{q(x-x')}L(x,x';q)$ is nothing but
the kernel $ L(x,x')$ of the original operator $\mathcal{L}(x,\partial _{x}^{}) $ in
(\ref{1}), we have from (\ref{14})
\begin{equation}
\fl \mathcal{L}(x,\partial
_{x}^{})\bigl(\rme_{}^{q(x-x')}M(x,x';q)\bigr)=\mathcal{L}_{}^{\mathrm{d}}(x',\partial
_{x'}^{})\bigl(\rme_{}^{q(x-x')}M(x,x';q)\bigr) =\delta (x-x'), \label{green}
\end{equation}
where $\mathcal{L}^{\mathrm{d}}$ is the operator dual to $\mathcal{L}$. Notice that while the
product $\rme^{q(x-x')}L(x,x';q)$ is $q$-independent the same combination for the resolvent
(see the simplest example in (\ref{M0xxq})) essentially depends on $q$. This means that the
resolvent can be considered as a two-parametric ($q\in \mathbb{R}^{2}$) family of Green's
functions of the operator $\mathcal{L}$.

Thanks to (\ref{251}) and (\ref{integralM}) the kernel $M(p;\mathbf{q})$, like
$M_{0}(p;\mathbf{q})$, is singular for $\mathbf{q}=\ell (\mathbf{q}_{1})$
and $\mathbf{q}+p=\ell (\mathbf{q}_{1}+p_{1})$, where we introduced the two
component vector
\begin{equation}
\ell (\alpha )=(\alpha ,-\rmi \alpha _{}^{2}),  \label{l}
\end{equation}
such that $L_{0}^{}(p;\ell (\mathbf{q}_{1}))\equiv 0$. A special role in the theory is played
by the operators $\nu $ and $\omega $, whose kernels in the ($p,\mathbf{q}$)-representation
are given as values of $(ML_{0})(p; \mathbf{q})$ and $(L_{0}^{}M)(p;\mathbf{q})$,
respectively, along these curves:
\begin{equation}
\fl \nu (p;\mathbf{q})=(ML_{0}^{})(p;\mathbf{q}) \Bigr|_{\mathbf{q}=\ell
(\mathbf{q}_{1})},\qquad \omega (p;\mathbf{q} )=(L_{0}^{}M)(p;\mathbf{q})
\Bigr|_{\mathbf{q}=\ell (\mathbf{q}_{1}+p_{1})-p}.  \label{181}
\end{equation}
It is clear by construction that both kernels $\nu (p;\mathbf{q})$ and $
\omega (p;\mathbf{q})$ are independent of $\mathbf{q}_{2}$ and it is easy to
see that both of them tend to $\delta (p)$ as $\mathbf{q}_{1}\rightarrow
\infty $. In the $x$-representation these kernels are given by means of the
inversion of (\ref{2a}) as
\begin{eqnarray}
\fl \nu (x,x';\mathbf{q}_{\Im }^{})=\frac{\delta (x_{2}^{}-x_{2}')}{2\pi }\!\!\int
\!\!\rmd\mathbf{q}_{1\Re }^{}\!\!\int\!\! \rmd x''\rme^{\rmi \ell _{\Re
}(\mathbf{q}_{1})(x'-x'')}(ML_{0}^{})(x,x'';\ell _{\Im }^{}
(\mathbf{q}_{1}^{})),  \label{nuxspace} \\
\fl \omega (x,x';\mathbf{q}_{\Im }^{})=\frac{\delta (x_{2}^{}-x_{2}')}{2\pi }\!\!\int \!\!\rmd
\mathbf{q}_{1\Re }^{}\!\!\int\!\! \rmd x''\rme^{\rmi \ell _{\Re
}(\mathbf{q}_{1})(x''-x)}\,(L_{0}^{}M)(x'',x';\ell _{\Im }^{}( \mathbf{q}_{1}^{})).
\label{omegaxspace}
\end{eqnarray}
Operators $\nu $ and $\omega $ obey the conjugation properties
\begin{equation}
\nu _{}^{\ast }=\nu ,\qquad \omega _{}^{\ast }=\omega , \label{conjugationnuomega}
\end{equation}
which are equivalent to (\ref{16}), and they are mutually inverse,
\begin{equation}
\omega \nu =I,\qquad \nu \omega =I.  \label{2326}
\end{equation}

The most essential property of these operators is that they dress $M_{0}$
and $L_{0}$ by the following formulae
\begin{equation}
M=\nu M_{0}\omega ,\qquad L=\nu L_{0}\omega ,  \label{25}
\end{equation}
that thanks to (\ref{2326}) give
\begin{equation}
\fl L\nu =\nu L_{0}^{},\qquad \omega L=L_{0}^{}\omega ,\qquad M\nu =\nu M_{0}^{},\qquad
\omega M=M_{0}^{}\omega .  \label{23}
\end{equation}
Thus it is natural to call $\nu $ and $\omega $ dressing operators, or more
specifically the Jost dressing operators, since the Jost solutions can be
directly related to them by means of the following construction. Let us
introduce
\begin{equation}
\Phi (x,\mathbf{k})=\rme_{}^{-\rmi \ell (\mathbf{k})x}\chi (x,\mathbf{k}),\qquad \Psi
(x,\mathbf{k})=\rme_{}^{\rmi \ell (\mathbf{k})x}\xi (x,\mathbf{k}), \label{3}
\end{equation}
where
\begin{equation}
\fl \chi (x,\mathbf{q}_{1}^{})=\!\!\int \!\!\rmd p\,\rme_{}^{-\rmi px} \nu
(p;\mathbf{q}),\qquad \xi (x,\mathbf{q}_{1}^{})=\!\!\int \!\!\rmd p\,\rme_{}^{-\rmi px}\omega
(p;\mathbf{q}-p) \label{183}
\end{equation}
and where we named $\mathbf{k}$ the spectral parameter $\mathbf{q}_{1}$ in
order to meet the traditional notation. Then the first pair of equalities in
(\ref{23}) takes the form
\begin{equation}
\mathcal{L}(x,\partial _{x}^{})\Phi (x,\mathbf{k})=0, \qquad
\mathcal{L}_{}^{\mathrm{d}}(x,\partial _{x}^{})\Psi (x,\mathbf{k})=0 \label{Schru}
\end{equation}
so that $\Phi $ and $\Psi $ solve the heat equation and its dual. In order
to prove that they are the Jost solutions we first note that for $\chi $ and
$\xi $ in (\ref{183}), thanks to (\ref{2a}) and (\ref{nuxspace}), (\ref
{omegaxspace}), we get
\begin{eqnarray}
\fl \chi (x,\mathbf{k}) =\!\!\int \!\!\rmd x'\,(ML_{0}^{})_{}^{(\ell _{\Re }(
\mathbf{k}))}(x,x';\ell _{\Im }^{}(\mathbf{k}))=\!\!\int \!\!\rmd x'\,\nu _{}^{(\mathbf{k})}
(x,x';0),  \label{d10} \\
\fl \xi (x',\mathbf{k}) =\!\!\int \!\!\rmd x\,(L_{0}^{}M)_{}^{(\ell _{\Re }(
\mathbf{k}))}(x,x';\ell _{\Im }^{}(\mathbf{k}))=\!\!\int \!\!\rmd x\,\omega
_{}^{(\mathbf{k})}(x,x';0).  \label{d10'}
\end{eqnarray}
The first equalities here can be used in (\ref{181}) and then we have by
(\ref{3}) and (\ref{183}) that
\begin{eqnarray}
\fl \Phi (x,\mathbf{k}) =\!\!\int \!\!\rmd x'\,\bigl(\mathcal{L}_{0}^{\mathrm{d} }(x',\partial
_{x'}^{})G(x,x',\mathbf{k})\bigr)
\rme_{}^{-\rmi \ell (\mathbf{k})x'},  \label{PhiG} \\
\fl \Psi (x',\mathbf{k}) =\!\!\int \!\!\rmd x\,\bigl(\mathcal{L}_{0}^{} (x,\partial
_{x}^{})G(x,x',\mathbf{k})\bigr) \rme_{}^{\rmi \ell (\mathbf{k})x},  \label{PsiG}
\end{eqnarray}
where the Green's function of the Jost solution
\begin{equation}
G(x,x',\mathbf{k})=\rme_{}^{q(x-x')}M(x,x';q) \Bigr|_{q=\ell _{\Im }(\mathbf{k})}^{}
\label{d1}
\end{equation}
appeared. Being a special case of the object considered in (\ref{green}) it
obeys
\begin{equation}
\mathcal{L}(x,\partial _{x}^{})G(x,x',\mathbf{k})= \mathcal{L}_{}^{\mathrm{d}}(x',\partial
_{x'}^{})G(x,x', \mathbf{k})=\delta (x-x'). \label{2305}
\end{equation}

From the integral equations (\ref{integralM}) we have $ ML_{0}^{}=I+M_{0}^{}UML_{0}^{}$ and $
L_{0}^{}M=I+L_{0}^{}MUM_{0}^{}$ that thanks to (\ref{PhiG}) and (\ref {PsiG}) give the
integral equations
\begin{eqnarray}
\Phi (x,\mathbf{k}) =\rme_{}^{-\rmi \ell (\mathbf{k})x}+\!\!\int \!\!\rmd
x'\,G_{0}^{}(x-x',\mathbf{k})u(x')\Phi (x',
\mathbf{k}),  \label{4} \\
\Psi (x,\mathbf{k}) =\rme_{}^{\rmi \ell (\mathbf{k})x}+\!\!\int \!\!\rmd
x'\,G_{0}^{}(x'-x,\mathbf{k})u(x')\Psi (x', \mathbf{k}), \label{4dual}
\end{eqnarray}
where the Green's function $G_{0}(x-x',\mathbf{k})$ is given by means of (\ref{d1}) in terms
of the bare resolvent $M_{0}$. It is easy to see that thanks to (\ref{M0xxq})
\begin{equation}
G_{0}(x,\mathbf{k})=\frac{1}{2\pi }\int \!\!\rmd \alpha \,\bigl[\theta (\mathbf{k}_{\Re }^{2}-
\alpha _{}^{2})-\theta (x_{2})\bigr]\,\rme_{}^{-\rmi \ell (\alpha +\rmi \mathbf{k}_{\Im })x},
\label{G0}
\end{equation}
which is the Green's function of the standard integral equations defining
the Jost solutions. Self-conjugation property of the operators $\nu $ and $
\omega $ (\ref{conjugationnuomega}) can be reformulated in terms of the Jost
solutions $\Phi (x,\mathbf{k})$ and $\Psi (x,\mathbf{k})$ in the following
way
\begin{equation}
\overline{\Phi (x,\mathbf{k})}=\Phi (x,-\overline{\mathbf{k}}),\qquad
\overline{\Psi (x,\mathbf{k})}=\Psi (x,-\overline{\mathbf{k}})
\end{equation}
as well as relations (\ref{2326}) that become the scalar product and
completeness relation for the Jost solutions
\begin{eqnarray}
\int \!\!\rmd x_{1}\Psi (x,\mathbf{k}+p)\Phi (x,\mathbf{k})=2\pi \delta (p),\quad
p\in \mathbb{R},  \label{xorthog} \\
\int\limits_{x_{2}'=x_{2}}\!\!\!\!\rmd\mathbf{k}_{\Re }\Psi (x', \mathbf{k})\Phi
(x,\mathbf{k})=2\pi \delta (x_{1}-x_{1}'). \label{xcompl}
\end{eqnarray}
Correspondingly, we call the first and the second relation in (\ref{2326})
the scalar product and the completeness relation for the dressing operators.

The dressing operators themselves can be reconstructed by means of the Jost
solutions,
\begin{eqnarray}
\fl \nu (x,x';q) =\frac{\delta (x_{2}^{}-x_{2}')}{2\pi }
\rme_{}^{-q_{1}(x_{1}-x_{1}')}\!\!\int \!\!\rmd k\,\,\rme_{}^{\rmi \ell
(k+\rmi q_{1})x'}\Phi (x,k+\rmi q_{1}^{}),  \label{d6} \\
\fl \omega (x,x';q) =\frac{\delta (x_{2}^{}-x_{2}')}{2\pi }
\rme_{}^{-q_{1}(x_{1}-x_{1}')}\!\!\int \!\!\rmd k\,\,\rme_{}^{-\rmi \ell (k+\rmi q_{1})x}\Psi
(x',k+\rmi q_{1}^{}).  \label{d7}
\end{eqnarray}
They can be expanded into the formal series
\begin{equation}
\nu =I+\sum_{n=1}^{\infty }\nu _{-n}^{}(2D_{1}^{})_{}^{-n},\qquad \omega
=I+\sum_{n=1}^{\infty }(2D_{1}^{})_{}^{-n}\omega _{-n}^{}, \label{exp:J}
\end{equation}
that in the ($p,\mathbf{q}$)-representation have the meaning of asymptotic
expansion in powers of $1/\mathbf{q}_{1}$ for $\nu (p;\mathbf{q})$ and of $
1/(\mathbf{q}_{1}+p_{1})$ for $\omega (p;\mathbf{q})$, i.e.\ the asymptotic
expansions of the Jost solutions with respect to the spectral parameter at
infinity. For the first coefficients in the expansion we get
\begin{equation}
\nu _{-1}+\omega _{-1}=0,\qquad \lbrack D_{1}^{},\nu _{-1}^{}]=-[D_{1}^{},\omega _{-1}^{}]=U,
\label{exp}
\end{equation}
that reconstruct the potential in the standard way, that is $u(x)=\partial
_{x_{1}}\nu _{-1}(x)$.

In order to formulate the Inverse problem we introduce the operator
\begin{equation}
\rho =L_{0}^{}ML_{0}^{}-L_{0}^{}.  \label{rho}
\end{equation}
that can be considered as a truncated resolvent. Then taking into account
the definition of the derivative in (\ref{2a10}) it is easy to show that the
$\bar{\partial}$-equations for the dressing operators are given in the form
\begin{equation}
\bar{\partial}_{1}^{}\nu =\nu R,\qquad \bar{\partial}_{1}^{}\omega =-R\omega ,  \label{26}
\end{equation}
where the operator $R$ is given by
\begin{equation}
R(p;\mathbf{q})=r(\mathbf{q}_{1}^{})\delta (p+2\ell _{\Re }^{} (\mathbf{q}_{1}^{})),
\label{27}
\end{equation}
with the spectral data $r(\mathbf{q}_{1})$ defined by means of the following
reduction of $\rho $:
\begin{equation}
r(\mathbf{q}_{1}^{})=2\pi\sgn (\mathbf{q}_{1\Re }^{})\rho (p;\ell (\mathbf{q}_{1}^{}))
\Bigr|_{p=-2\ell _{\Re }^{}(\mathbf{q}_{1}^{})}^{}.  \label{271}
\end{equation}
Let us mention that by construction the operator $R$ obeys the conditions
\begin{eqnarray}
\lbrack L_{0}^{},R] =0,\qquad \lbrack M_{0}^{},R]=0,  \label{28} \\
R^{\ast } =-R.  \label{29}
\end{eqnarray}
In terms of the Jost solutions equations (\ref{26}) take the standard form
\begin{equation}
\fl \overline{\partial }_{\mathbf{k}}\Phi (x,\mathbf{k})=\Phi (x,-\overline{
\mathbf{k}})r(\mathbf{k}),\qquad \overline{\partial }_{\mathbf{k}}\Psi (x,
\mathbf{k})=-\Psi (x,-\overline{\mathbf{k}})r(-\overline{\mathbf{k}}),
\label{2308}
\end{equation}
where $\overline{r(\mathbf{k})}=-r(-\overline{\mathbf{k}})$ (cf. (\ref{29})).
In order to get the representation of the spectral data in terms of the
Jost solutions we use (\ref{12}) and (\ref{14}) and rewrite (\ref{rho}) in
the two equivalent forms $\rho =UML_{0}$ and $\rho =L_{0}MU$. Then for the
value of $\rho $ in (\ref{rho}) we get by (\ref{181}), (\ref{3}) and (\ref
{183}) that
\begin{equation}
\fl r(\mathbf{k})=\frac{\sgn \mathbf{k}_{\Re }^{}}{2\pi }\!\!\int \!\!\rmd x\,\rme_{}^{\rmi
\ell (-\bar{\mathbf{k}} )x}u(x)\Phi (x,\mathbf{k})= \frac{\sgn \mathbf{k}_{\Re }^{}}{2\pi
}\!\!\int \!\!\rmd x\,\rme_{}^{-\rmi \ell (\mathbf{k})x}u(x)\Psi (x,-\bar{\mathbf{k}}).
\label{rk}
\end{equation}
Finally, the normalization conditions that complete the formulation of the
inverse problem for the Jost solution or its dual are given by (\ref{3}) in
the standard forms $\lim_{\mathbf{k}\rightarrow \infty }\chi (x,\mathbf{k})=1
$ or, correspondingly, $\lim_{\mathbf{k}\rightarrow \infty }\xi (x,\mathbf{k}
)=1$ as follows from (\ref{183}) and (\ref{exp:J}).

\section{Similarity transformations of the Spectral Data}

\subsection{Rational similarity transformations}

We study transformations of the above introduced objects generated by
similarity transformations of the spectral data of the form
\begin{equation}
R'=WRW_{}^{-1}  \label{31}
\end{equation}
where the operator $W$ is the following rational function of the operator $
D_{1}$:
\begin{equation}
W=\prod_{j=1}^{N}\frac{D_{1}^{}-a_{j}^{}}{D_{1}^{}-b_{j}^{}}, \label{701}
\end{equation}
with $a_{j}$ and $b_{j}$ some complex parameters. Such $W$ guarantees that $ R'$ is of the
form (\ref{27}) with
\begin{equation}
r'(\mathbf{k})=\left( \prod_{j=1}^{N}\frac{\mathbf{k}-\rmi b_{j}^{}}{ \mathbf{\bar{k}}+\rmi
b_{j}^{}}\frac{\mathbf{\bar{k}}+\rmi a_{j}^{}}{\mathbf{k} -\rmi a_{j}^{}}\right) r(\mathbf{k})
\label{528}
\end{equation}
substituted for $r$. It is easy to see that such $R'$ obeys properties (\ref{28}), i.e.\ it
commutes with $L_{0}$ and $M_{0}$, and in order for $R'$ to obey property (\ref{29}) we have
to impose the following conditions on the parameters:
\begin{equation}
a_{j}^{}=\bar{a}_{\pi _{a}(j)}^{},\qquad b_{j}^{}=\bar{b}_{\pi _{b}(j)}^{}  \label{702}
\end{equation}
where $\pi _{a}$ and $\pi _{b}$ are some permutations of the indices. Notice that in the
simplest situation where $N=1$, i.e.\ $ W=(D_{1}-a_{1})/(D_{1}-b_{1})$, the parameters must be
real, $a_{1}=\bar{a}_{1}$, $b_{1}=\bar{b}_{1}$. Like in the nonstationary Schr\"{o}dinger case
the potential $u'$ corresponding to this simplest situation can be obtained from the
potential $u$ by means of a binary Darboux transformation suggested in \cite{matveev}.
However, in contrast with the nonstationary Schr\"{o}dinger equation, the generic case, with
$W$ given by (\ref{701}) and $a_{j}^{}$, $b_{j}^{}$ subjected to conditions (\ref{702}) but
not necessarily real, cannot be obtained by applying recursively binary Darboux
transformations or one needs to admit non real intermediate potentials in the iterative
procedure. Here we shall not use the techniques of the Darboux transformations but rather
derive these transformations solving the inverse problem given by the spectral data $R'$
explicitly in terms of the objects corresponding to the generic original spectral data $R$.

Let us mention that the new spectral data $R'$ have additional discontinuities if compared
with $R$. Indeed, $r'(\mathbf{k})$ given in (\ref{528}) is undetermined at points
$\mathbf{k}=\rmi a_{j}$ and $ \mathbf{k}=\rmi b_{j}$ while $r(\mathbf{k})$ as given in
(\ref{rk}) is a well defined function for $\mathbf{k}_{\Re }\neq 0$. Correspondingly, these
additional discontinuities lead to new properties for the new dressing operators $\nu'$ and
$\omega '$. Indeed, we show below the solvability of the $\bar{\partial}$-equations
\begin{equation}
\fl \bar{\partial}_{1}^{}\nu '=\nu 'R'+\rmi \pi \sum_{j=1}^{N}\nu _{a_{j}}'\delta
(D_{1}^{}-a_{j}^{}),\qquad \bar{\partial}_{1}^{}\omega '=-R'\omega '+\rmi \pi
\sum_{j=1}^{N}\delta (D_{1}^{}-b_{j}^{})\omega _{b_{j}}', \label{720}
\end{equation}
that differ from equations (\ref{26}) by additional $\delta $-terms at $ D_{1}=a_{j}$ and at
$D_{1}=b_{j}$. In other words we show that $\nu'$ and $\omega '$ can have poles on the complex
plane with residua $ \nu _{a_{j}}'$ and $\omega _{b_{j}}'$. Different choices
of the form of these additional terms correspond to some renormalizations of
the dressing operators and, consequently, of the Jost solutions\footnote[7]{
In \cite{Barbara}, where the direct problem was examined, in order to define
Jost solutions via an integral equation invariant in form, i.e.\ not depending
on the parameters $a_{j}$ and $b_{j}$, a different normalization was
necessary. This is another intriguing characteristic of the heat equation in
comparison with the nonstationary Schr\"{o}dinger equation.}. One could also
consider the case in which poles are absent, but we will show that, then, the
solution does not contain solitons and decays at large distances.

In order to close the formulation of the inverse problem for $\nu '$ and $\omega '$ we have to
impose normalization conditions on $\nu '(p;\mathbf{q})$ and $\omega'(p;\mathbf{q})$ at some
value of $\mathbf{q}_{1}$ (both of them, as we know, must be independent on the variable
$\mathbf{q}_{2}$). Taking into account that $\nu $ and $\omega $ are normalized by the
asymptotic condition at infinity, it is natural to choose the same point for the $\nu '$ and
$\omega '$. Then it is easy to see that, without loss of generality (omitting the
uninteresting case of a potential $u(x)$ shifted by a function of $x_{2}$ only), we can fix
that
\begin{equation}
\nu '(p;\mathbf{q)}\rightarrow \delta (p),\qquad \omega '(p;\mathbf{q)}\rightarrow \delta
(p),\quad \mathbf{q}_{1}\rightarrow \infty .  \label{370}
\end{equation}

In this discussion we used that $\bar{\partial}_{1}(D_{1}-a)^{-1}=\rmi \pi \delta
(D_{1}^{}-a)$. The kernels of this $\delta $-operator in the $x$- and
($p,\mathbf{q}$)-representations are given by
\begin{eqnarray}
\bigl(\delta (D_{1}^{}-a)\bigr)(p;\mathbf{q}) =\delta
(\mathbf{q}_{1}^{}-\rmi a)\delta (p),  \nonumber \\
\bigl(\delta (D_{1}^{}-a)\bigr)(x,x';q) =\frac{\rme_{}^{\rmi a_{\Im }(x_{1}-x_{1}')}}{2\pi
}\delta (q_{1}^{}-a_{\Re }^{})\delta (x_{2}^{}-x_{2}'), \label{708}
\end{eqnarray}
where we used the definition $\delta (z)=\delta (z_{\Re })\delta (z_{\Im })$ for any $z\in
\mathbb{C}$. Note that in the ($p,\mathbf{q}$)-representation $ \nu _{}'(p;\mathbf{q})$ and
$\omega '(p;\mathbf{q})$ have poles, respectively, at $\mathbf{q}_{1}=\rmi a_{j}$ and
$\mathbf{q}_{1}=\rmi b_{j}-p_{1}$ and we have
\begin{equation}
\fl \nu _{a_{j}}'(p;\mathbf{q})=-\rmi \res_{\mathbf{q}_{1}= \rmi a_{j}}
\nu'(p;\mathbf{q}_{1}^{}),\qquad \omega _{b_{j}}'(p;\mathbf{q})=-\rmi \res_{\mathbf{q}_{1}=
\rmi b_{j}-p_{1}}\omega '(p;\mathbf{q}_{1}^{}).  \label{721}
\end{equation}
Correspondingly, the Jost solutions $\Phi '(x,\mathbf{k})$ and $ \Psi'(x,\mathbf{k})$ have
poles, respectively, at $\mathbf{k} =\rmi a_{j}$ and $\mathbf{k}=\rmi b_{j}$ and in the
$x$-representation we have
\begin{eqnarray}
\nu _{a_{j}}'(x,x';q) =\delta (x-x')\rme_{}^{\rmi \ell (\rmi a_{j})x}\Phi _{a_{j}}'(x),  \\
\omega _{b_{j}}'(x,x';q) =\delta (x-x')\rme_{}^{-\rmi \ell (\rmi b_{j})x'}\Psi _{b_{j}}'(x')
\label{nuaomegabxx'}
\end{eqnarray}
where
\begin{equation}
\Phi _{a_{j}}'(x)=-\rmi \res_{\mathbf{k}=\rmi a_{j}}\Phi _{}'(x,\mathbf{k}),\qquad \Psi
_{b_{j}}'(x)=-\rmi \res_{\mathbf{k}=\rmi b_{j}}\Psi _{}'(x,\mathbf{k}). \label{35'}
\end{equation}

Note that in the formulation of the inverse problem the values of the residua at poles $\nu
_{a_{j}}'$ and $\omega _{b_{j}}'$ are left free. We show below that in order to close the
construction of $\nu '$ and $\omega '$ it is necessary to impose that they dress the $L_{0}$
operator, that is, that they obey the differential equations like the first pair of equations
in (\ref{23})
\begin{equation}
L'\nu '=\nu 'L_{0}^{},\qquad \omega 'L'=L_{0}^{}\omega ', \label{36}
\end{equation}
where
\begin{equation}
L'=L_{0}^{}-U'  \label{37}
\end{equation}
with some new potential $u'$. More precisely, we show in Section 3.2 that if $\nu '$ and
$\omega '$ are dressing operators they satisfy the orthogonality relation $\omega'\nu '=I$
and, then, in Section 3.3, that this additional requirement imposed to $\nu '$ and $\omega '$
obtained from the solution of the $ \overline{\partial }$ problem (\ref{720}) is sufficient in
order to guarantee that they satisfy the dressing equations (\ref{36}). If poles are absent
$\nu '$ and $\omega '$ are uniquely determined by (\ref{720}) and their asymptotic behaviour
at large $\mathbf{q}$ and it is not necessary to impose any additional requirement in order
that they satisfy equations (\ref{36}).

Let us note that the new dressing operators, since they obey (\ref{36}), have the same
(formal) expansions with respect to $\mathbf{q}_{1}$ as $\nu $ and $\omega $ in (\ref{exp:J})
with new $\nu _{-n}'$, $\omega _{-n}'$ substituted for $\nu _{-n}$, $\omega _{-n}$.

We also assume that the new potential is real, i.e.\ $U'{}^{\ast }=U'$ and, correspondingly,
$L'{}^{\ast }=L'$. Moreover, we assume that there exists the resolvent $M'$ of the new
operator $L'$, that is its inverse operator. So we have also the second pair of equations in
(\ref{23})
\begin{equation}
M'\nu '=\nu 'M_{0}^{},\qquad \omega 'M'=M_{0}^{}\omega '. \label{360}
\end{equation}

In order to solve the $\bar{\partial}$-equations for the new dressing
operators in terms of the old ones it is convenient to consider, first, the $
\bar{\partial}$-equations
\begin{equation}
\bar{\partial}_{1}^{}(\nu'W\omega )=\nu' (\bar{\partial}_{1}^{}W)\omega ,\qquad
\bar{\partial}_{1}^{}(\nu W_{}^{-1}\omega')=\nu (\bar{\partial}_{1}^{}W_{}^{-1})\omega',
\label{35}
\end{equation}
that can be obtained from (\ref{26}), (\ref{31}) and (\ref{720}) noting that
$W$ and $W^{-1}$ cancel exactly the additional $\delta $ terms in (\ref{720}).
In order to integrate explicitly these equations we write first that by (\ref{701})
\begin{equation}
\fl \bar{\partial}_{1}^{}W=\rmi \pi \sum_{j=1}^{N}c_{j}^{}\,\delta
(D_{1}^{}-b_{j}^{}),\qquad \bar{\partial}_{1}^{}W_{}^{-1}=\rmi \pi \sum_{j=1}^{N}
\tilde{c}_{j}^{}\,\delta (D_{1}^{}-a_{j}^{})
\end{equation}
where
\begin{equation}
c_{j}^{}=\frac{\displaystyle\prod_{l=1}^{N}(b_{j}^{}-a_{l})}{{\displaystyle
\mathop{{\prod\nolimits'}}_{l=1}^{N}}(b_{j}^{}-b_{l})},
\qquad \tilde{c}_{j}^{}=\frac{\displaystyle
\prod_{l=1}^{N}(a_{j}^{}-b_{l})}{{\displaystyle
\mathop{{\prod\nolimits'}}_{l=1}^{N}}(a_{j}^{}-a_{l}^{})}  \label{707}
\end{equation}
and $^{\prime}$ means that the term $l=j$ is omitted. Thus (\ref{35}) takes
the form
\begin{eqnarray}
\bar{\partial}_{1}^{}(\nu'W\omega )=\rmi \pi \sum_{j=1}^{N}c_{j}^{}\nu_{b_{j}}' \delta
(D_{1}^{}-b_{j}^{})\omega _{b_{j}}^{},\label{710}
\\ \bar{\partial}_{1}^{}(\nu W_{}^{-1}\omega')= \rmi \pi
\sum_{j=1}^{N}\tilde{c}_{j}^{}\nu _{a_{j}}^{}\delta (D_{1}^{}-a_{j}^{})\omega _{a_{j}}'
\label{7101}
\end{eqnarray}
where new operators were introduced whose kernels in the
($p$,$\mathbf{q}$)-representation are independent on $\mathbf{q}$ and are,
precisely, values
of the dressing operators at some points
\begin{eqnarray}
\nu _{b}'(p;\mathbf{q}) =\nu'(p;\rmi b),\qquad \omega
_{a}'(p;\mathbf{q})=\omega'(p;\rmi a-p_{1}^{}),  \label{711} \\
\nu _{a}^{}(p;\mathbf{q}) =\nu (p;\rmi a),\qquad \omega _{b}^{}(p;\mathbf{q}
)=\omega (p;\rmi b-p_{1}^{}).  \label{712}
\end{eqnarray}
Thanks to the composition law (\ref{180}) and (\ref{708}) it is easy to see
that $\nu _{a}^{}\delta (D_{1}^{}-a)=\nu \delta (D_{1}^{}-a)$, $\delta
(D_{1}^{}-b)\omega _{b}^{}=\delta (D_{1}^{}-b)\omega $, etc. Just these
equalities were used in (\ref{710}) and (\ref{7101}) where in the r.h.s.'s thanks to them
only $\delta $-functions have kernels that in ($p$,$\mathbf{q}$)-representation
depend on $\mathbf{q}_{1}$. This means that we can rewrite
(\ref{710}) and (\ref{7101}) as
\begin{eqnarray}
\bar{\partial}_{1}^{}\left( \nu'W\omega -\sum_{j=1}^{N}\nu
_{b_{j}}'\frac{c_{j}^{}}{D_{1}^{}-b_{j}^{}}\omega _{b_{j}}^{}\right)
=0,\\
\bar{\partial}_{1}^{}\left( \nu W_{}^{-1}\omega'-\sum_{j=1}^{N}\nu
_{a_{j}}^{}\frac{\tilde{c}_{j}^{} }{D_{1}^{}-a_{j}^{}}\omega _{a_{j}}'\right) =0.
\end{eqnarray}
Since the kernels of $\nu'$, $\nu $, $\omega'$, $\omega $ and $W$ in the
($p$,$\mathbf{q}$)-representation tend to $\delta (p)$ when $ \mathbf{q}_{1}\rightarrow \infty
$, the expressions in parenthesis tend to $ I $ in the same limit and the equations can be
explicitly integrated. Taking into account (\ref{2326}) we have
\begin{eqnarray}
\nu'=\left( I+\sum_{j=1}^{N}\nu _{b_{j}}'\frac{c_{j}^{}}{
D_{1}^{}-b_{j}^{}}\omega _{b_{j}}^{}\right) \nu W_{}^{-1},\label{718}\\
\omega'=W\omega \left( I+\sum_{j=1}^{N}\nu _{a_{j}}^{}
\frac{\tilde{c}_{j}^{}}{D_{1}^{}-a_{j}^{}}\omega _{a_{j}}'\right)\label{7181}.
\end{eqnarray}

\subsection{Scalar products of the dressing operators}

\label{scalarproduct}Representations (\ref{718}) and (\ref{7181}) obtained in the previous
subsection are still undetermined, since they include the unknown multiplication operators
$\nu _{b_{j}}'$ and $\omega _{a_{j}}'$. In order to define them we need, first, to evaluate
the scalar product of the dressing operators
\begin{equation}
S=\omega'\nu'.  \label{751}
\end{equation}
Directly from this definition and thanks to (\ref{720}) we have
\begin{equation}
\bar{\partial}_{1}^{}S=[S,R']+\rmi \pi \sum_{j=1}^{N}S_{a_{j}}^{}\delta
(D_{1}^{}-a_{j}^{})+\rmi \pi \sum_{j=1}^{N}\delta (D_{1}^{}-b_{j}^{})S_{b_{j}}^{},
\label{752}
\end{equation}
where we used notation (\ref{721}) for the residuum at a pole. In getting, for instance, the
second term in the r.h.s.\ we used the relation $ \omega'\nu _{a}'\delta
(D_{1}^{}-a)=\omega'(\nu'(D_{1}^{}-a))\delta (D_{1}^{}-a)=(S(D_{1}^{}-a))\delta
(D_{1}^{}-a)=S_{a}\delta (D_{1}^{}-a)$, which is obtained by noting that $\omega'$ can be
inserted into the bracket since it has no pole in $a$.

Thanks to (\ref{36}) and (\ref{360}) $S$ must commute with $L_{0}$ and $ M_{0} $. The
condition of commutativity of $S$ with $M_{0}$ formally is a consequence of the commutativity
with $L_{0}$ as their are inverse one to another. However, as we already mentioned, the
composition law (\ref {180xspace}) is not necessarly associative and therefore both conditions
must be imposed. Let us consider $[L_{0},S]=0$ in the ($p$,$ \mathbf{q}$)-representation. We
obtain from (\ref{180}) and (\ref{13}) the equation $ \left[ \rmi
p_{2}^{}-p_{1}^{}(p_{1}^{}+2\mathbf{q}_{1}^{})\right] S(p; \mathbf{q})=0$. Then
$S(p;\mathbf{q})$ must be a generalized function concentrated at $p=0$ and at $p+2\ell _{\Re
}(\mathbf{q}_{1})=0$ and, consequently, a linear combination with arbitrary coefficients
depending on $ \mathbf{q}_{1}$ of $\delta \left( p\right) $, $\delta \left( p+2\ell _{\Re }(
\mathbf{q}_{1})\right) $ and their derivatives up to a finite order. The derivative to be
considered is $\left( \frac{\partial }{\partial p_{1}^{}} -2\rmi
(p_{1}^{}+\mathbf{q}_{1}^{})\frac{\partial }{\partial p_{2}^{}} \right) $ since it must
annihilate $\left[ \rmi p_{2}^{}-p_{1}^{}(p_{1}^{}+2\mathbf{q}_{1}^{})\right] $. This
derivative when applied to $\delta \left( p+2\ell _{\Re } (\mathbf{q}_{1})\right) $ can be
equivalently written as $\frac{\partial }{\partial \mathbf{\bar{q}}_{1}^{}}$ and, therefore,
we conclude that the most general $S$ commuting with $L_{0}$ has a kernel which is a finite
linear combination (with coefficients depending on $\mathbf{q}_{1}$) of the following
distributions
\begin{equation}
\fl Y_{n}^{}(p;\mathbf{q})=\left[ \frac{\partial }{\partial p_{1}^{}} -2\rmi
(p_{1}^{}+\mathbf{q}_{1}^{})\frac{\partial }{\partial p_{2}^{}} \right] _{}^{n}\delta
(p),\quad Z_{n}(p;\mathbf{q})=\frac{\partial ^{n}}{
\partial \mathbf{\bar{q}}_{1}^{n}}\delta (p+2\ell _{\Re }(\mathbf{q}_{1})),
\end{equation}
where $n=0,1,\dots $. Let us consider now the condition $[M_{0},S]=0$. One
can check that only $Y_{0}$ and $Z_{0}$ commute with $M_{0}$, while $Y_{n}$
and $Z_{n}$ for $n=1,2,\ldots $ do not commute with it and, precisely, the
most singular terms in $[M_{0},Y_{n}]$ and in $[M_{0},Z_{n}]$ are,
respectively, proportional to $\frac{\partial ^{n-1}}{\partial \mathbf{\bar{q
}}_{1}^{n-1}}\delta \left( \rmi \mathbf{q}_{2}-\mathbf{q}_{1}^{2}\right) \delta
(p)$ and $\frac{\partial ^{n-1}}{\partial \mathbf{\bar{q}}_{1}^{n-1}}\delta
\left( \rmi \mathbf{q}_{2}-\mathbf{q}_{1}^{2}\right) \delta (p+2\ell _{\Re }(
\mathbf{q}_{1}))$, which, of course, does not contradict the statement that $
Y_{n}$ and $Z_{n}$ commute with $L_{0}$. Therefore $S$ can only be a linear
combination of $Y_{0}$ and $Z_{0}$. But the term with $Z_{0}$ substituted in
the l.h.s.\ of (\ref{752}) produces a term $Z_{1}$ that cannot be
compensated by a term in the r.h.s.

Thus finally we conclude that kernel of the operator $S$ in the ($p$,$
\mathbf{q}$)-representation must be of the form
\begin{equation}
S(p;\mathbf{q})=s(\mathbf{q}_{1})\delta (p).  \label{768}
\end{equation}
Now turning back to the equation (\ref{752}) we see that in the ($p$,$
\mathbf{q}$)-representation the term in the l.h.s.\ as well as the second
and third terms in the r.h.s.\ are proportional to $\delta (p)$ while the
first term in the r.h.s.\ is proportional to $\delta (p+2\ell _{\Re }(
\mathbf{q}_{1}))$ due to (\ref{27}). So this term has to be equal to zero:
\begin{equation}
\lbrack S,R'](p;\mathbf{q})\equiv (s(-\mathbf{\bar{q}}_{1}^{})-s(
\mathbf{q}_{1}^{}))r'(\mathbf{q}_{1}^{})\delta (p+2\ell _{\Re }^{}( \mathbf{q}_{1}^{}))=0
\label{529}
\end{equation}
and (\ref{752}) is reduced to
\begin{equation}
\bar{\partial}_{1}^{}S=\rmi \pi \sum_{j=1}^{N}S_{a_{j}}^{}\delta
(D_{1}^{}-a_{j}^{})+\rmi \pi \sum_{j=1}^{N}\delta
(D_{1}^{}-b_{j}^{})S_{b_{j}}^{}.
\end{equation}
Taking into account that thanks to (\ref{370}) and (\ref{751}) at
$\mathbf{q}_{1}$-infinity $S(p;\mathbf{q})\rightarrow \delta (p)$ we get
\begin{equation}
S=I+\sum_{j=1}^{N}\left( \frac{S_{a_{j}}^{}}{D_{1}^{}-a_{j}^{}}+\frac{
S_{b_{j}}^{}}{D_{1}^{}-b_{j}^{}}\right)
\end{equation}
or by (\ref{768})
\begin{equation}
s(\mathbf{q}_{1}^{})=1+\sum_{j=1}^{N}\left(
\frac{\rmi S_{a_{j}}^{}}{\mathbf{q}_{1}^{}-\rmi a_{j}^{}}+
\frac{\rmi S_{b_{j}}^{}}{\mathbf{q}_{1}^{}-\rmi b_{j}^{}}\right) .
\end{equation}

Finally, it is clear that for generic $r(\mathbf{q}_{1})$ (and, thus, $ r'(\mathbf{q}_{1})$)
equation (\ref{529}) gives that $ S_{a_{j}}=S_{b_{j}}=0$ for all $j$ and thus we proved that
\begin{equation}
S=\omega'\nu'=I.  \label{538}
\end{equation}
In other words we proved that like in the case of decaying potential the dressing operators
obey the first equality in (\ref{2326}). It is necessary to mention that, on the contrary, as
shown below, the second equality for the product $\nu'\omega'$ changes essentially.

\subsection{Construction of Jost solutions and potential}

In order to proceed with the construction of the dressing operators $\nu '$ and $\omega '$ it
is convenient to work in the $x$-representation and to introduce the corresponding solutions
$\Phi '(x,\mathbf{k})$ and $\Psi '(x,\mathbf{k})$ by means of the $ '$ analog of the
relations (\ref{3}) and (\ref{183}). Let us first consider the object
$f_{b}^{}=(D_{1}-b)^{-1}\omega _{b}\nu $ appearing in the expression for $\nu '$ in
(\ref{718}). By (\ref{101}), (\ref {183}) and (\ref{3}) we have
\begin{equation}
\fl f_{b}^{}(x,x';q)=\frac{\delta (x_{2}^{}-x_{2}')}{2\pi } \!\!\int \!\!\rmd \alpha
\,\rme_{}^{\rmi (\ell (\alpha +\rmi q_{1})-\ell (\rmi b))x+\rmi \alpha
(x_{1}'-x_{1})}\mathcal{F}(x,\rmi b,\alpha +\rmi q_{1}^{})  \label{727}
\end{equation}
where the so called Cauchy--Baker--Akhiezer function introduced in \cite{GO}
\begin{equation}
\mathcal{F}(x,\mathbf{k},\mathbf{k}')=\int\limits^{x_{1}}_{ (\mathbf{k}_{\Im
}-\mathbf{k}_{\Im }')\infty \atop y_2=x_2 }\!\!\!\!\rmd y_{1}^{}\,\Psi (y,\mathbf{k})\Phi
(y,\mathbf{k}') \label{729}
\end{equation}
had appeared. Then by using (\ref{d10}) we obtain
\begin{equation}
\fl \Phi '(x,\mathbf{k})=\prod_{l=1}^{N}\frac{\mathbf{k}-\rmi b_{l}^{}}{ \mathbf{k}-\rmi
a_{l}^{}}\left( \Phi (x,\mathbf{k})+\sum_{j=1}^{N}c_{j}^{} \Phi '(x,\rmi
b_{j}^{})\mathcal{F}(x,\rmi b_{j}^{},\mathbf{k})\right) . \label{7490}
\end{equation}
Analogously from $\omega '$ in (\ref{7181}) we have
\begin{equation}
\fl \Psi '(x,\mathbf{k})=\prod_{l=1}^{N}\frac{\mathbf{k}-\rmi a_{l}^{}}{ \mathbf{k}-\rmi
b_{l}^{}}\left( \Psi (x,\mathbf{k})-\sum_{j=1}^{N}
\tilde{c}_{j}^{}\mathcal{F}(x,\mathbf{k},\rmi a_{j}^{})\Psi '(x,\rmi a_{j}^{})\right) .
\label{7491}
\end{equation}
The function $\mathcal{F}(x,\mathbf{k},\mathbf{k}')$ obeys the $ \bar{\partial}$-equations
\begin{eqnarray}
\frac{\partial }{\partial \mathbf{\bar{k}}'}\mathcal{F}(x,\mathbf{k} ,\mathbf{k}') =\rmi \pi
\delta (\mathbf{k}-\mathbf{k}')+ \mathcal{F}(x,\mathbf{k},-\mathbf{\bar{k}}')r(\mathbf{k}'),
\label{734} \\
\frac{\partial }{\partial \mathbf{\bar{k}}}\mathcal{F}(x,\mathbf{k},\mathbf{k }') =-\rmi \pi
\delta (\mathbf{k}-\mathbf{k}')-\mathcal{F}
(x,-\mathbf{\bar{k}},\mathbf{k}')r(-\mathbf{\bar{k}}),  \label{735}
\end{eqnarray}
that are often considered as a generalization of equations (\ref{2308}).

If $\Phi '(x,\mathbf{k})$ and $\Psi '(x,\mathbf{k})$ have no pole singularities, i.e.\ $\Phi
_{a_{j}}'(x)\equiv \Psi _{b_{j}}'(x)\equiv 0$ (see (\ref{35'})) for all $j$, then the
denominators in the r.h.s.'s of (\ref{7490}) and (\ref{7491}) must be compensated by zeroes of
the expressions in parentheses. These $2N$ equations uniquely determine the $2N$ functions
$\Phi '(x,\rmi b_{j}^{})$ and $\Psi '(x,\rmi a_{j}^{})$ and then the Jost solutions $\Phi
'(x,\mathbf{k})$ and $\Psi '(x,\mathbf{k})$ themselves. In this case it is easy to check that
the corresponding dressing operators obey (\ref{36}) that proves the statement given there for
the absence of poles. Turning back to the generic situation, in order to complete the
construction of $\Phi '(x,\mathbf{k})$ and $\Psi '(x,\mathbf{k})$ we evaluate the values $\Phi
'(x,\rmi b_{j}^{})$ and $\Psi '(x,\rmi a_{j}^{})$ appearing in (\ref {7490}) and (\ref{7491})
by using the equality $\omega '\nu '=I$ derived in the previous section. Once constructed
$\Phi '(x,\mathbf{k})$ and $\Psi '(x,\mathbf{k})$ in terms of $ \Phi (x,\mathbf{k})$ and $\Psi
(x,\mathbf{k})$, by using the asymptotic limits $\lim_{x_{2}\rightarrow \infty }\chi
(x,\mathbf{k})=\lim_{x_{2} \rightarrow \infty }\xi (x,\mathbf{k})=1$, easily derived from
(\ref{4}) and (\ref{4dual}), we must verify that $\chi '(x,\mathbf{k})$ and
$\xi'(x,\mathbf{k})$ are polynomially bounded at space infinity and, then, that $\nu '$ and
$\omega '$ belong to the considered space of operators, which guarantees the correctness of
the above derived equalities and, in particular, of (\ref{538}). We assume this behavior and
below we shall prove it in some special cases.

It was already mentioned that the first equality in (\ref{2326}) can be rewritten in terms of
$\Phi $ and $\Psi $ as (\ref{xorthog}). The same orthogonality condition for $\Phi '$ and
$\Psi '$ follows from (\ref{538}). Inserting in it (\ref{7490}) and (\ref{7491}) we get
\begin{eqnarray}
\fl \int \!\!\rmd x_{1}^{} \left( \Psi (x,\mathbf{k}+p)-\sum_{j=1}^{N}
\tilde{c}_{j}^{}\mathcal{F}(x,\mathbf{k}+p,\rmi a_{j}^{})\Psi '(x,\rmi a_{j}^{})\right)
\nonumber \\
\times  \left( \Phi (x,\mathbf{k})+\sum_{j=1}^{N}c_{j}^{}\Phi '(x,\rmi
b_{j}^{})\mathcal{F}(x,\rmi b_{j}^{},\mathbf{k})\right) =2\pi \delta (p).
\end{eqnarray}
Thanks to the above assumption it is clear that the integral exists in the sense of
distributions. Now we multiply this equality by $\Psi (x', \mathbf{k})\Phi (x'',\mathbf{k}+p)$
where $x_{2}'=x_{2}''=x_{2}$, integrate it with respect to $\mathbf{k}_{\Re }$ and $p$ and use
(\ref{xcompl}). Then the r.h.s.\ becomes equal to $ 4\pi ^{2}\delta (x_{1}'-x_{1}'')$.
Correspondingly, the l.h.s.\ at a generic point $\mathbf{k}$ must have zero $\partial _{
\overline{\mathbf{k}}}$-derivative and the residua of poles at $\mathbf{k} =\rmi a_{j}$ and
$\mathbf{k}=\rmi b_{j}$ must be equated to zero getting
\begin{eqnarray}
\Psi '(x,\rmi a_{j}^{})=\sum_{l=1}^{N}\mathcal{F}'(x,\rmi a_{j}^{},\rmi b_{l}^{})c_{l}^{}
\Psi (x,\rmi b_{l}^{}),  \label{521} \\
\Phi '(x,\rmi b_{j}^{})=-\sum_{l=1}^{N}\Phi (x,\rmi a_{l}^{})
\tilde{c}_{l}^{}\mathcal{F}'(x,\rmi a_{l}^{},\rmi b_{j}^{}),  \label{522}
\end{eqnarray}
where in analogy with (\ref{729}) we introduced
\begin{equation}
\mathcal{F}'(x,\mathbf{k},\mathbf{k}')= \int\limits^{x_{1}}_{(\mathbf{k}_{\Im }
-\mathbf{k}_{\Im }')\infty \atop y_{2}=x_{2}} \!\!\!\!\rmd
y_{1}^{}\,\Psi'(y,\mathbf{k})\Phi'(y,\mathbf{k}'). \label{517}
\end{equation}
Multiplying (\ref{521}) by (\ref{522}) we get
\begin{eqnarray}
\fl \Psi '(x,\rmi a_{j}^{})\Phi '(x,\rmi b_{k}^{})\nonumber
\\=-\sum_{l,m=1}^{N}\mathcal{F}'(x,\rmi a_{j}^{},\rmi b_{l}^{})c_{l}^{}\Psi
(x,\rmi b_{l}^{})\Phi (x,\rmi a_{m}^{})\tilde{c}_{m}^{}\mathcal{F}'(x,\rmi a_{m}^{},\rmi
b_{k}^{}).  \label{523}
\end{eqnarray}
Now by (\ref{517}) we rewrite this as
\begin{eqnarray}
\fl \frac{\partial }{\partial x_{1}^{}}\mathcal{F}'(x,\rmi a_{j}^{},\rmi b_{k}^{})\nonumber
\\=-\sum_{l,m=1}^{N}\mathcal{F}'(x,\rmi a_{j}^{},\rmi b_{l}^{})c_{l}^{}\left(
\frac{\partial }{\partial x_{1}^{}}\mathcal{F}(x,\rmi b_{l}^{},\rmi a_{m}^{})\right)
\tilde{c}_{m}^{} \mathcal{F}'(x,\rmi a_{m}^{},\rmi b_{k}^{}),
\end{eqnarray}
or as
\begin{equation}
\frac{\partial }{\partial x_{1}^{}}(\mathcal{F}')_{}^{-1}=\frac{
\partial }{\partial x_{1}^{}}c\mathcal{F}\tilde{c},
\end{equation}
where we introduced matrices $\mathcal{F}$ and $\mathcal{F}'$ with elements
\begin{equation}
\mathcal{F}_{jk}^{}=\mathcal{F}(x,\rmi b_{j},\rmi a_{k}),\qquad
\mathcal{F}_{jk}'=\mathcal{F}'(x,\rmi a_{j},\rmi b_{k})  \label{matrix}
\end{equation}
and the two diagonal matrices $c=\mathrm{diag}\{c_{1},\ldots c_{N}\}$ and $
\tilde{c}=\mathrm{diag}\{\tilde{c}_{1},\ldots \tilde{c}_{N}\}$. Let $C$
denote an $x_{1}$-independent matrix. Thus we get
\begin{equation}
\mathcal{F}'=\tilde{c}_{}^{-1}(C+\mathcal{F})_{}^{-1}c_{}^{-1} \label{5270}
\end{equation}
and inserting $\mathcal{F}'$ in (\ref{521}) and (\ref{522}), correspondingly,
\begin{eqnarray}
\Psi '(x,\rmi a_{j}^{}) =\frac{1}{\tilde{c}_{j}^{}}
\sum_{l=1}^{N}(C+\mathcal{F})_{jl}^{-1}\Psi (x,\rmi b_{l}^{}),  \label{5271} \\
\Phi '(x,\rmi b_{j}^{}) =\frac{-1}{c_{j}^{}}\sum_{l=1}^{N}\Phi (x,\rmi
a_{l}^{})(C+\mathcal{F})_{lj}^{-1}.  \label{5272}
\end{eqnarray}
In order to avoid singularities we have to impose the condition that $\det
(C+\mathcal{F}(x))$ has no zeroes on the $x$-plane. At the end of this Section we show that in
the cases $N=1$ and $N=2$ this condition is formulated as a condition on admissible matrices
$C$. For generic $N$ we assume the existence of such matrices $C$ that guarantee absence of
zeroes of this determinant. In addition one has to check that for these $C$ the matrix
$\mathcal{F}'$ with elements $\mathcal{F}_{jk}'$ obtained from (\ref{517}) by taking
$\mathbf{k}=\rmi a_{j}$ and $\mathbf{k}'=\rmi b_{k}$ and by substituting, respectively,
(\ref{5271}) and (\ref {5272}) for $\Phi '(x,\rmi b_{k})$ and $\Psi '(x,\rmi a_{j})$
coincides with (\ref{5270}). This leads to the equality
\begin{equation}
\mathcal{F}_{jk}'(x)=\int\limits_{(a_{j}-b_{k})\infty }^{x_{1}}\!\!\rmd
y_{1}^{}\frac{\partial }{\partial y_{1}^{}} \mathcal{F}_{jk}'(y_{1}^{},x_{2}^{}),
\end{equation}
i.e.\ all elements of this matrix given by (\ref{5270}) have to obey
conditions
\begin{equation}
\lim_{x_{1}^{}\rightarrow (a_{j}-b_{k})\infty }\mathcal{F}_{jk}'(x)=0.  \label{condf}
\end{equation}
It is easy to check these conditions in the cases $N=1$ and $N=2$, but for
generic $N$ we have to assume their validity.

Now from (\ref{7490}) and (\ref{7491}) we get for the residua of $\Phi '$ and $\Psi'$
\begin{equation}
\fl \Phi _{a_{j}}'(x)=-\tilde{c}_{j}^{}\sum_{l=1}^{N}\Phi '(x,\rmi
b_{l}^{})c_{l}^{}C_{lj}^{},\qquad \Psi
_{b_{j}}'(x)=c_{j}^{}\sum_{l=1}^{N}C_{jl}^{}\tilde{c}_{l}^{}\Psi '(x,\rmi a_{l}^{})
\label{5273}
\end{equation}
We can also insert the r.h.s.'s of (\ref{5271}) and (\ref{5272}) into (\ref
{7490}) and (\ref{7491}) getting
\begin{eqnarray}
\fl \Phi '(x,\mathbf{k}) =\prod_{m=1}^{N}\frac{\mathbf{k}-\rmi b_{m}^{} }{\mathbf{k}-\rmi
a_{m}^{}}\left( \Phi (x,\mathbf{k})-\sum_{j,l=1}^{N}\Phi (x,\rmi
a_{j}^{})(C+\mathcal{F})_{jl}^{-1}\mathcal{F}(x,\rmi b_{l}^{},\mathbf{k}
)\right) ,  \label{5274} \\
\fl \Psi '(x,\mathbf{k}) =\prod_{m=1}^{N}\frac{\mathbf{k}-\rmi a_{m}^{} }{\mathbf{k}-\rmi
b_{m}^{}}\left( \Psi (x,\mathbf{k})-\sum_{j,l=1}^{N}\mathcal{ F}(x,\mathbf{k},\rmi
a_{j}^{})(C+\mathcal{F})_{jl}^{-1}\Psi (x,\rmi b_{l}^{})\right) . \label{5275}
\end{eqnarray}
The above formulas can be rewritten in terms of ratio of determinants as it was done for the
KPI-case \cite{Steklov}. They can be considered as a special case of the construction given in
\cite{matveev} for generic eigenfunctions, i.e.\ not necessarily Jost solutions. On the
contrary, here we deal with objects that indeed can be called Jost solutions as their
departure from analyticity is under control and follows from (\ref{720}) by the $'$ analog of
the relations (\ref{3}) and (\ref{183}) and equalities (\ref{nuaomegabxx'}):
\begin{eqnarray}
\overline{\partial }_{\mathbf{k}}\Phi '(x,\mathbf{k})
=\Phi'(x,-\overline{\mathbf{k}})r'(\mathbf{k})+\rmi\pi \sum_{j=1}^{N}\Phi _{a_{j}}'(x)\delta
(\mathbf{k}-\rmi a_{j}^{}),
\\
\overline{\partial }_{\mathbf{k}}\Psi '(x,\mathbf{k})
=-\Psi'(x,-\overline{\mathbf{k}})r'(-\overline{\mathbf{k}})+\rmi \pi \sum_{j=1}^{N}\Psi
_{b_{j}}'(x)\delta (\mathbf{k}-\rmi b_{j}^{}), \label{5276}
\end{eqnarray}
where $\Phi _{a_{j}}'(x)$ and $\Psi _{b_{j}}'(x)$ obey (\ref{5273}). Thanks to (\ref{35'}) and
(\ref{5274}), (\ref{5275}) we get explicitly
\begin{eqnarray}
\Phi _{a_{j}}'(x) =\tilde{c}_{j}^{}\sum_{l,m=1}^{N}\Phi
(x,\rmi a_{l}^{})(C+\mathcal{F})_{lm}^{-1}C_{mj}^{},  \label{5277} \\
\Psi _{b_{j}}'(x) =c_{j}^{}\sum_{l,m=1}^{N}C_{jl}^{}(C+ \mathcal{F})_{lm}^{-1}\Psi (x,\rmi
b_{m}^{}),  \label{5278}
\end{eqnarray}
where we used notations (\ref{707}) and (\ref{matrix}). In particular, the above discussed
case where $\Phi '(x,\mathbf{k})$ and $\Psi '(x,\mathbf{k})$ have no poles corresponds to
matrix $C=0$. These transformations do not add solitons and can be viewed as transformations
of the continuous spectrum.

Now we can derive an explicit formula for the potential $u'$ with spectral data (\ref{528}).
Inserting in (\ref{5274}) the two leading terms of the asymptotic expansion (\ref{exp:J}) and
(\ref{exp:J})$'$, i.e.\ $\chi (x,\mathbf{k})=1-(2\rmi \mathbf{k})^{-1}\chi _{-1}(x)+\ldots $
and its analog for $\chi'$, we get thanks to (\ref{3})
\begin{equation}
\chi _{-1}'(x)=\chi _{-1}^{}(x)+2\sum_{j=1}^{N}(a_{j}^{}-b_{j}^{})-2\partial _{x_{1}}^{}\log
\det (C+\mathcal{F}(x)),\qquad
\end{equation}
so that for the potential we have
\begin{equation}
u'(x)=u(x)-2\partial _{x_{1}}^{2}\log \det (C+\mathcal{F}(x)). \label{452}
\end{equation}

It is easy to check directly that if $\mathcal{L}'(x,\partial _{x})$ denote the operator of
the type (\ref{1}) with potential $u'(x)$ substituted for $u(x)$ then (\ref{5274}) and
(\ref{5275}) obey equations $ \mathcal{L}'(x,\partial _{x})\Phi'(x,\mathbf{k})=0$ and its dual
if and only if
\begin{equation}
\partial _{x_{2}}^{}C=0.  \label{453}
\end{equation}
Formula (\ref{452}) is well known in the literature \cite
{generalKPsolitons,matveev}, but nevertheless conditions that guarantee the
regularity of the new potential are, to our knowledge, absent. This
situation is completely different from the case of the Nonstationary
Schr\"{o}dinger Equation, where conditions of regularity were given in \cite
{Steklov}. If all $a_{j}$ and $b_{j}$ in (\ref{701}) are real, it is enough
to choose the matrix $C$ real in order to get a real potential. However, in
the generic situation when (\ref{702}) is satisfied, the conditions to
impose to $C$ in order to have a real potential are unknown. Even more
complicated are the requirements to impose to the matrix $C$ in order to
obtain regular potentials.

We made a complete analysis in the cases $N=1$ and $N=2$ when the $a_{j}$ and $b_{j}$ are
real. Then the matrix $C$ must be real. In the case $N=1$ the matrix $C$ reduces to a real
constant and the regularity condition is $ (a_{1}-b_{1})C\geq 0$. The case $C=0$ corresponds
to a non solitonic situation and, in particular, when the original potential $u(x)$ is equal
to zero we get that $u'(x)=0$ also. In the case $N=2$ a complete analysis is elementary but
rather lengthy. Here, we give only the final result, more details are presented in the
Appendix. It is convenient to introduce the following constants
\begin{equation}
D_{ij}=C_{ij}J_{i+1,j+1},\qquad J_{ij}=\frac{1}{a_{j}-b_{i}},
\end{equation}
where $i$ and $j$ are defined $\mod 2$ and to choose, for definiteness,
$a_{1}<a_{2}$ and $b_{1}<b_{2}$.

In order to have a potential describing two solitons superimposed to a
background at least two $D_{ij}$ must be different from zero.

Then the potential is regular and has four rays if and only if for $i=1,2$

\begin{description}
\item[in the case $\det J<0$, $J_{11}J_{22}>0$]\makebox{}
\begin{equation}
D_{ii}\leq 0,\qquad D_{i,i+1}>0,\qquad \det C\leq 0
\end{equation}
\item[in the case $\det J<0$, $J_{11}J_{22}<0$]\makebox{}
\begin{equation}
D_{ii}<0,\qquad D_{i,i+1}\geq 0,\qquad \det C\leq 0
\end{equation}
\item[in the case $\det J>0$]\makebox{}
\begin{equation}
D_{ii}\geq 0,\qquad D_{i,i+1}<0.
\end{equation}
or
\begin{equation}
D_{ii}>0,\qquad D_{i,i+1}\leq 0.
\end{equation}
\end{description}
The potential is regular and has three rays if and only if for $i=1$ or $2$
and $j=1$ or $2$
\begin{description}
\item[in the case $\det J<0$]\makebox{}
\begin{equation}
D_{ii}=D_{j,j+1}=0,\qquad D_{i+1,i+1}<0,\qquad D_{j+1,j}>0
\end{equation}
\item[in the case $\det J>0$]\makebox{}
\begin{equation}
D_{ii}=D_{j,j+1}=0,\qquad D_{i+1,i+1}>0,\qquad D_{j+1,j}<0.
\end{equation}
\end{description}

If three $D_{ij}$ are zero the potential contains only one soliton.

Notice that the two soliton solution $u'(x)$ can be considered as a surface in the $(x,u')$
space depending on $8$ parameters, i.e. $ D_{ij}$ ($i,j=1,2$), $a_{i}$ and $b_{i}$ ($i=1,2$).
Since in the space of parameters the surfaces $D_{ij}=0$ and $\det C=0$ separate singular from
regular solutions the soliton solution is not differentiable on these surfaces, which are
therefore bifurcation surfaces according to the usual definition in catastrophe theory.
Therefore, we expect that the solution, as a geometrical object, would be structurally
instable at these values of the parameters.

The possible behaviours of the potential at large distances in the
$x$-plane, according to the different choices of the parameters, are richer
than in the KPI case. For instance, in the case $\det J>0$, if we call $
x_{1}+hx_{2}=$ const. the direction of a soliton, the directions at $
x_{2}=\pm \infty $ of the two solitons when the matrix $C$ is full are given
by $h=a_{1}+a_{2}$ and $h=b_{1}+b_{2}$, when the matrix $C$ is diagonal by $
h=a_{1}+b_{1}$ and $h=a_{2}+b_{2}$ and when the matrix $C$ is off-diagonal
by $h=a_{1}+b_{2}$ and $h=a_{2}+b_{1}$. In addition if only one element of
the matrix $C$ is zero the four rays of the solitons are directed along four
different directions. These different behaviours are in agreement with the
previous comment on the structural stability of the soliton solution. In the
appendix we give a detailed description of the potentials in all regular
cases.

\section{Completeness relation and resolvent}

We proved in section \ref{scalarproduct} that, like in the case of decaying potentials, the
scalar product (\ref{538}) $\omega '\nu '$ is equal to $I$ or, in other words, that $\nu '$ is
right inverse of $\omega '$. On the contrary we expect, due to the presence of discrete data
in the spectrum, that the second equality in (\ref{2326}), the so called completeness
relation, is modified by an additional operator $ P'$ as follows
\begin{equation}
\nu '\omega '+P'=I.  \label{457}
\end{equation}
From (\ref{538}) we get directly that
\begin{equation}
P'\nu '=0,\qquad \omega 'P'=0,\qquad P_{}^{\prime 2}=P', \label{459}
\end{equation}
i.e.\ $P'$ is an orthogonal projector. By substituting in $\nu '\omega '$ the values of $\nu
'$ and $\omega '$ given in (\ref{718}) and (\ref{7181}) we get an expression of $P'$ in terms
of the residua of $\nu '$ and $\omega '$
\begin{equation}
P'=-\sum_{j=1}^{N}\nu _{a_{j}}'\frac{I}{ D_{1}^{}-a_{j}^{}}\omega _{a_{j}}'-\sum_{j=1}^{N}\nu
_{b_{j}}'\frac{I}{D_{1}^{}-b_{j}^{}}\omega _{b_{j}}' \label{460}
\end{equation}
that in the $x$-representation can be written explicitly as
\begin{eqnarray}
\fl P'(x,x';q) =-\sgn(x_{1}^{}-x_{1}')\delta
(x_{2}^{}-x_{2}')\rme_{}^{-q_{1}(x_{1}-x_{1}')}   \nonumber \\
\times \sum_{j=1}^{N}\{\theta ((q_{1}^{}-b_{j\Re }^{})(x_{1}^{}-x_{1}'))\Phi '(x,\rmi
b_{j}^{})\Psi_{b_{j}}'(x')  \nonumber \\
+\theta ((q_{1}^{}-a_{j\Re }^{})(x_{1}^{}-x_{1}'))\Phi _{a_{j}}'(x)\Psi '(x',\rmi
a_{j}^{})\}.  \label{461}
\end{eqnarray}
This equality can be simplified by means of (\ref{5273})
\begin{eqnarray}
\fl P'(x,x';q)=-\delta (x_{2}^{}-x_{2}')\rme_{}^{-q_{1}(x_{1}-x_{1}')}   \nonumber \\
\times \sum_{j,l=1}^{N}\Phi '(x,\rmi b_{j}^{})c_{j}^{}C_{jl}^{}\tilde{c}_{l}^{}\Psi'(x',\rmi
a_{l}^{})\{\theta (q_{1}^{}-b_{j\Re }^{})-\theta (q_{1}^{}-a_{l\Re }^{})\}. \label{462}
\end{eqnarray}
Then we see that $P'$ is equal to zero outside the largest interval in $q_{1}$ with extremes
$a_{l\Re }^{}$, $b_{j\Re }^{}$ ($j,l=1,\ldots ,N $) and that $\overline{\partial }P'$
(see (\ref{2a12})) is equal to zero if $q_{1}$ is different from $a_{l\Re }^{}$ and $b_{j\Re
}^{}$. Again it is necessary to emphasize that all these results are obtained on the basis of
the assumption of polynomially boundedness of the $\chi '$ and $\xi '$ formulated in the
previous section.

Also the dressing formula for the resolvent $M'$, inverse of $ L'$, must be corrected with
respect to the bilinear representation (\ref{25}) adding an operator $m'$ as follows
\begin{equation}
M'=\nu'M_{0}\omega'+m'.  \label{463}
\end{equation}
Thanks to (\ref{36}) we have $L'M'=L'\nu^{ \prime}M_{0}\omega'+L'm'=\nu '\omega'+L'm'$, where
we used the fact that $ L_{0}M_{0}=I$. Thus in order to obey $L'M'=I$ the operator $ m'$ by
(\ref{457}) has to obey the equality $L'm^{ \prime}=P'$.

Let us consider here the case $N=1$ and let $a=a_{1}$, $b=b_{1}$ be real and
let us label the corresponding quantities with $1$. We have then
\begin{equation}
M_{1}^{}=\nu _{1}^{}M_{0}^{}\omega _{1}^{}+m_{1}^{}  \label{463-1}
\end{equation}
and, in this case, it is easy to derive from (\ref{460}) that
\begin{eqnarray}
\fl m_{1}^{}= \nu _{1,a}^{}\frac{1}{
(D_{1}^{}-a)(D_{2}^{}-(a+b)D_{1}^{}+ab)}\omega _{1,a}^{}  \nonumber \\
+\nu _{1,b}^{}\frac{1}{(D_{1}^{}-b)(D_{2}^{}-(a+b)D_{1}^{}+ab)} \omega _{1,b}^{}. \label{464}
\end{eqnarray}
In writing the kernel of this operator in the $x$-representation we can use
(\ref{5273}) that in this case thanks to (\ref{707}) takes the form
\begin{equation}
\Phi _{1,a}^{}(x)=c\Phi _{1}^{}(x,\rmi b)\qquad \Psi _{1,b}^{}(x)=-c\Psi _{1}^{}(x,\rmi a),
\label{mfirst}
\end{equation}
where
\begin{equation}
c=(a-b)_{}^{2}C_{11}^{}.
\end{equation}
We get a formula similar to (\ref{462}), that is
\begin{eqnarray}
\fl m_{1}^{}(x,x';q)= \bigl(\theta (q_{1}^{}-a)-\theta (q_{1}^{}-b)\bigr)\bigl[\theta
(x_{2}^{}-x_{2}')-\theta
(-q_{2}^{}+(a+b)q_{1}^{}-ab)\bigr]  \nonumber \\
\times c\rme_{}^{-q(x-x')}\Phi _{1}^{}(x,ib)\Psi _{1}^{}(x',\rmi a). \label{msecond}
\end{eqnarray}
From this equation we have $L_{1}m_{1}=P_{1}$ as required. Let us, however, stress that we
would obtain this equality for any other $x$-independent second term in the square bracket and
that the above specific choice is necessary in order to avoid an exponential growth of
$m_{1}(x,x';q)$ at space infinities, as easily follows from relations (\ref{5271}) and (\ref
{5272}). For generic $N$ the construction of $m'$ and the study of its asymptotic properties
are essentially more cumbersome.

Let us discuss here some properties of the resolvent $M_{1}$ constructed for the case $N=1$ as
they follow from (\ref{463-1}) and (\ref{464})--(\ref {msecond}). We see that, in comparison
with the resolvent $M$ (\ref{25}) of a decaying potential, $M_{1}$ has an additional
discontinuity at $ q_{2}=(a+b)q_{1}-ab$ due to the term $m_{1}( x,x';q)$. This discontinuity
is not compensated by the first term in (\ref{463-1}) and its presence in the resolvent
$M_{1}$ is a characteristic manifestation of the solitonic content of the potential
$u_{1}(x)$. The term $m_{1}( x,x';q) $ is zero when $q_{1}$ is outside the interval $( a,b) $
and therefore discontinuous also along the lines $q_{1}=a$ and $q_{1}=b$ on the $ q$-plane.
These discontinuities can be compensated by the pole behavior of the dressing operators $\nu
_{1}$ and $\omega _{1}$ in the dressed term $\nu _{1}M_{0}\omega _{1}$ and need a detailed
study.

From (\ref{720}) using the Cauchy-Green formula and the asymptotic behaviour of $\nu _{1}$ and
$\omega _{1}$ we have
\begin{eqnarray}
\nu _{1}^{} =I-\frac{1}{\pi }\int \frac{\rmd_{}^{2}\mathbf{z}_{1}^{}}{
\mathbf{z}_{1}^{}}\,\bigl(\nu _{1}^{}R_{1}^{}\bigr)_{}^{(
\mathbf{z}_{1}^{})}+\nu _{1,a}^{}\frac{1}{D_{1}^{}-a}, \\
\omega _{1}^{} =I+\frac{1}{\pi }\int \frac{\rmd_{}^{2}\mathbf{z}_{1}^{}
}{\mathbf{z}_{1}^{}}\,\bigl(R_{1}^{}
\omega_{1})^{(\mathbf{z}_{1})}+\frac{1}{D_{1}^{}-b}\omega _{1,b}^{}
\end{eqnarray}
where notation (\ref{2a10}) for the shift was used. Then inserting them into
$\nu _{1}M_{0}\omega _{1}$ we get
\begin{equation}
\nu _{1}^{}M_{0}^{}\omega _{1}^{}=M_{0}^{}-\frac{1}{\pi }\int \frac{
\rmd_{}^{2}\mathbf{z}_{1}^{}}{\mathbf{z}_{1}^{}}\nu _{1}^{(
\mathbf{z}_{1})}\bigl[R_{1}^{(\mathbf{z}_{1})},M_{0}^{}\bigr] \omega
_{1}^{(\mathbf{z}_{1})}+M_{1,\mathrm{discr}}^{},
\end{equation}
where
\begin{equation}
M_{1,\mathrm{discr}}^{}=\nu _{1,a}^{}\frac{M_{0}^{}}{D_{1}^{}-a}\omega _{1,a}^{}+\nu
_{1,b}^{}\frac{M_{0}^{}}{D_{1}-b}\omega _{1,b}^{}.
\end{equation}
In the $x$-representation we get for the contribution of the discrete part
of the spectrum to $\nu _{1}M_{0}\omega _{1}$, thanks to (\ref{101}), (\ref
{M0xxq}), (\ref{nuaomegabxx'}), and (\ref{5273}),
\begin{equation}
\fl M_{1,\mathrm{discr}}^{}(x,x';q)=c\Phi _{1}^{}(x,\rmi b)\Psi _{1}^{}(x',\rmi
a)\rme_{}^{-q(x-x')}[\Gamma _{a}(x-x';q)-\Gamma _{b}(x-x';q)],
\end{equation}
where
\begin{equation}
\Gamma _{a}^{}(x;q)=\frac{\rme_{}^{\rmi\ell (\rmi a)x}}{2\pi }\!\!\int \!\! \rmd \alpha
\bigl[ \theta (q_{1}^{2}-q_{2}^{}-\alpha _{}^{2})-\theta (x_{2}^{})\bigr]
\frac{\rme_{}^{-\rmi\ell (\alpha +\rmi q_{1})x}}{q_{1}^{}-a-\rmi\alpha }.
\end{equation}
The singularities of $\Gamma _{a}$ can be explicitly extracted getting
\begin{equation}
\Gamma _{a}^{}(x;q)=\Gamma _{a,\mathrm{reg}}^{}(x;q)+\Gamma _{a,\mathrm{sing} }^{}(x;q),
\end{equation}
where
\begin{eqnarray}
\fl \Gamma _{a,\mathrm{reg}}^{}(x;q)= \frac{\theta (x_{2})}{2}\left\{ 1-\erf
\frac{x_{1}+2ax_{2}}{2\sqrt{x_{2}^{}}}\right\}   \nonumber \\
-\frac{\theta (q_{1}^{2}-q_{2}^{})}{2\pi \rmi}\int\limits_{-\sqrt{
q_{1}^{2}-q_{2}}}^{\sqrt{q_{1}^{2}-q_{2}}}\!\!\rmd\alpha \frac{\rme_{}^{-\rmi \ell (\alpha
+\rmi q_{1})x+\rmi \ell (\rmi a)x}-1}{\alpha +\rmi (q_{1}^{}-a)}
\end{eqnarray}
and
\begin{equation}
\Gamma _{a,\mathrm{sing}}^{}(x;q)=-\theta (x_{2})\theta (q_{1}^{}-a)+\frac{ \theta
(q_{1}^{2}-q_{2}^{})}{\pi }\arctan \frac{\sqrt{q_{1}^{2}-q_{2}^{}} }{q_{1}^{}-a}.
\end{equation}
We conclude that, for generic $q_{2}$, the discontinuities along the lines $ q_{1}=a$ and
$q_{1}=b$ of $M_{1,\mathrm{discr}}(x,x';q)$ cancel exactly the discontinuities along the same
lines of $m_{1}$, but $ \rme^{-qx}\Gamma _{a,\mathrm{sing}}(x;q)$ and $\rme^{-qx}\Gamma
_{b,\mathrm{sing} }(x;q) $ in the neighborhood, respectively, of the points $(a,a^{2})$ and $
(b,b^{2})$ in the $q$-plane are ill defined due to the $\arctan $.

One of the main advantages in using the resolvent is that the Green's
functions can be obtained as specific reduction with respect to the
parameter $q$ of the resolvent. In particular, if we are interested in
considering a smooth perturbation $u_{2}(x)$ (decaying at space infinity) of
the potential $u_{1}(x)$ constructed above, then the Jost solution $
\widetilde{\Phi }(x,\mathbf{k})$ of the perturbed potential $\tilde{u}
=u_{1}+u_{2}$ can be obtained as a perturbation of the Jost solution $\Phi
_{1}(x,\mathbf{k})$ by means of the following integral equation
\begin{equation}
\widetilde{\Phi }(x,\mathbf{k})=\Phi _{1}^{}(x,\mathbf{k})+\!\!\int\!\! \rmd x'\,
G_{1}^{}(x,x',\mathbf{k})u_{2}^{}(x') \widetilde{\Phi}(x', \mathbf{k}), \label{Jpert}
\end{equation}
where (cf.\ (\ref{d1})) $G_{1}(x,x',\mathbf{k})=\rme^{q(x-x^{ \prime})}M_{1}(x,x';q)\bigr|
_{q=\ell _{\Im}(\mathbf{k})}$. Taking into account that thanks to (\ref{l})
$\ell_{\Im}(\mathbf{k})= (\mathbf{k}_{\Im },-\mathbf{k}_{\Re }^{2}+\mathbf{k}_{\Im }^{2})$ we
get that the above mentioned discontinuity coming from $m_{1}$ becomes a discontinuity across
the hyperbole $\left( \mathbf{k}_{\Im }-\frac{a+b}{2}\right) ^{2}- \mathbf{k}_{\Re }^{2}=
\left( \frac{a-b}{2}\right) ^{2}$ which lies in the $ \mathbf{k}$-plane outside the strip
$a<\mathbf{k}_{\Im }<b$ if $a<b$ or $b< \mathbf{k}_{\Im }<a$ if $b<a$. On the other side the
term in $G_{1}$ coming from $m_{1} $ is equal to zero outside this strip. Therefore, according
to our previous discussion, only the discontinuities at $\mathbf{k}=\rmi a$ and at $
\mathbf{k}=\rmi b$ are left. Thus in the case of the heat equation (at least for $N=1$) the
Green's function of the Jost solution has no an additional cuts in contrast with the case of
the nonstationary Schr\"{o}dinger equation \cite {FokasPogreb,towards}, but, anyway, due to
the special singularities at $ \mathbf{k}=\rmi a$ and at $\mathbf{k}=\rmi b$, the definition
of the spectral data also for the case of a perturbed one soliton potential is not standard
and needs a detailed analysis. This problem will be faced in a following work.

\ack A.K.P. acknowledges financial support from INFN
and thanks colleagues at Lecce Department of Physics for kind hospitality
and fruitful discussions.

\appendix

\section*{Appendix}
\setcounter{section}{1} The potential $u'(x) $ describing two solitons superimposed to a
generic background has at large distances in the $x$-plane a solitonic one dimensional
behaviour along some rays. Any specific such ray can be represented by an equation
$x_{1}+hx_{2}=$ const. with a given $h$ and by specifying if the rays is pointing to
$x_{2}=+\infty $ or to $ x_{2}=-\infty .$ We choose, for definiteness,
$a_{2}>a_{1}$ and $b_{2}>b_{1}$.

Let us, first, list the cases in which the potentials has four rays.

It is convenient to rename the parameters $a_{i}$ and $b_{i}$ ($i=1,2$) as $
\alpha _{j}$ ($j=1,\dots ,4)$ in such a way that
\begin{equation}
\alpha _{1}<\alpha _{2}<\alpha _{3}<\alpha _{4}.
\end{equation}
If $C_{ij}\neq 0$ and $\det C\neq 0$ we have the following rays
\begin{equation}
\begin{tabular}{|l|l|}
\hline
h & versus \\ \hline
$\alpha _{1}+\alpha _{3}$ & $\pm \infty $ \\ \hline
$\alpha _{2}+\alpha _{4}$ & $\pm \infty $ \\ \hline
\end{tabular}
.  \label{fullC}
\end{equation}
Note that the rays are two by two parallel.

According to the remark we made at the end of Section 3 we expect that when
an element of the matrix $C$ or $\det C$ is taken to be zero the asymptotic
behaviour at large distances will change discontinuously. If only one
element of the matrix $C$ or $\det C$ is zero we have four rays at large
distances but they are no longer two by two parallel. If two elements not
belonging to the same diagonal of $C$ are zero we are left with only three
rays. If three elements are zero the solution reduces to the one soliton
solution and if the matrix $C$ is zero the solution does not contain
solitons.

The details are the following.

If, in the space of parameters of the soliton solution, we move to a
bifurcation point by taking $C_{ii}=0$ for $i=1$ and/or $2$ ($\det C\neq 0$)
the four rays are obtained by transforming discontinuously among the rays
listed in the previous table (\ref{fullC}) the following rays for $i=1$
and/or $2$
\begin{equation}
\begin{tabular}{|l|l|}
\hline
h & versus \\ \hline
$\alpha _{1}+\alpha _{3}$ & $\left( -\right) ^{i+1}J_{11}\infty $ \\ \hline
$\alpha _{2}+\alpha _{4}$ & $\left( -\right) ^{i+1}J_{11}\infty $ \\ \hline
\end{tabular}
\end{equation}
into the rays
\begin{equation}
\begin{tabular}{|l|l|}
\hline
h & versus \\ \hline
$\alpha _{1}+\alpha _{4}$ & $\left( -\right) ^{i+1}J_{11}\infty $ \\ \hline
$\alpha _{2}+\alpha _{3}$ & $\left( -\right) ^{i+1}J_{11}\infty $ \\ \hline
\end{tabular}
.
\end{equation}
If $C_{i,i+1}=0$ for $i=1$ and/or $2$ and $\det C\neq 0$ the four rays are
obtained by substituting for $i=1$ and/or $2$ the following rays
\begin{equation}
\begin{tabular}{|l|l|}
\hline
h & versus \\ \hline
$\alpha _{1}+\alpha _{3}$ & $\left( -\right) ^{i}\infty $ \\ \hline
$\alpha _{2}+\alpha _{4}$ & $\left( -\right) ^{i+1}\infty $ \\ \hline
\end{tabular}
\end{equation}
among the rays listed in the table (\ref{fullC}) with the rays
\begin{equation}
\begin{tabular}{|l|l|}
\hline
h & versus \\ \hline
$\alpha _{1}+\alpha _{2}$ & $\left( -\right) ^{i}\infty $ \\ \hline
$\alpha _{3}+\alpha _{4}$ & $\left( -\right) ^{i+1}\infty $ \\ \hline
\end{tabular}
.
\end{equation}
If $\det C=0$, $\det J<0$ and $J_{11}J_{22}>0$ and all $C_{ij}\neq 0$ the
four rays are
\begin{equation}
\begin{tabular}{|l|l|}
\hline
h & versus \\ \hline
$\alpha _{1}+\alpha _{2}$ & $-J_{11}\infty $ \\ \hline
$\alpha _{1}+\alpha _{3}$ & $+J_{11}\infty $ \\ \hline
$\alpha _{2}+\alpha _{4}$ & $-J_{11}\infty $ \\ \hline
$\alpha _{3}+\alpha _{4}$ & $+J_{11}\infty $ \\ \hline
\end{tabular}
.
\end{equation}
If $\det C=0$, $\det J<0$ and $J_{11}J_{22}<0$ and all $C_{ij}\neq 0$ the
four rays are
\begin{equation}
\begin{tabular}{|l|l|}
\hline
h & versus \\ \hline
$\alpha _{1}+\alpha _{3}$ & $-J_{11}\infty $ \\ \hline
$\alpha _{1}+\alpha _{4}$ & $+J_{11}\infty $ \\ \hline
$\alpha _{2}+\alpha _{3}$ & $+J_{11}\infty $ \\ \hline
$\alpha _{2}+\alpha _{4}$ & $-J_{11}\infty $ \\ \hline
\end{tabular}
.
\end{equation}

Let us, now, list the cases in which the potential has three rays.

If $C_{ii}=0$ and $C_{i,i+1}=0$ for $i=1$ or $2$ the three rays are
\begin{equation}
\begin{tabular}{|l|l|}
\hline
h & versus \\ \hline
$a_{1}+a_{2}$ & $+J_{i+1,i+1}J_{i+1,i}\infty $ \\ \hline
$a_{i}+b_{i+1}$ & $\left( -\right) ^{i+1}J_{i+1,i+1}\infty $ \\ \hline
$a_{i+1}+b_{i+1}$ & $\left( -\right) ^{i}J_{i+1,i}\infty $ \\ \hline
\end{tabular}
.
\end{equation}

If $C_{ii}=0$ and $C_{i+1,i}=0$ for $i=1$ or $2$ the three rays are
\begin{equation}
\begin{tabular}{|l|l|}
\hline
h & versus \\ \hline
$b_{1}+b_{2}$ & $-J_{i+1,i+1}J_{i,i+1}\infty $ \\ \hline
$a_{i+1}+b_{i}$ & $\left( -\right) ^{i+1}J_{i+1,i+1}\infty $ \\ \hline
$a_{i+1}+b_{i+1}$ & $\left( -\right) ^{i}J_{i,i+1}\infty $ \\ \hline
\end{tabular}
.
\end{equation}
Finally, let us note that the corrections to this one dimensional solitonic
behaviour are exponentially decaying, in contrast with the KPI case, where
the corrections are rationally decaying at least in some regions of the
plane \cite{towards}.

\end{document}